\documentclass[aps,preprint,amsmath,amssymb,showpacs,amsfonts]{revtex4}
\usepackage{epsfig}
\usepackage{graphicx}
\usepackage{subfigure}
\usepackage{dcolumn}
\usepackage{bm}

\newcommand{\SLTC}{SL(2, \mathbb{C})}
\newcommand{\tr}{\operatorname{tr}}

\begin{document}

\title{Parity and reality properties of the EPRL spinfoam}

\author{Yasha Neiman}
\email{yashula@gmail.com}
\affiliation{Raymond and Beverly Sackler School of Physics and Astronomy, Tel-Aviv University, Tel-Aviv 69978, Israel}

\date{\today}

\begin{abstract}
We study the parity behavior of the Lorentzian EPRL spinfoam model. We demonstrate that the vertex amplitude does not depend on the sign of the Immirzi parameter. We present numerical results for the transition amplitude and the graviton propagator in the large-spin 4-simplex approximation. The results suggest a simple relation between the contributions of the two parity-related critical points. Finally, we observe that the graviton propagator is not invariant under parity-odd permutations of equivalent nodes. Thus, the Lorentzian model has the same chirality problem as the Euclidean.
\end{abstract}
\maketitle

\section{Introduction}

In recent years, the covariant approach to Lorentzian loop quantum gravity has converged on a single model - the ``new'' spinfoams \cite{FK,EPRL}. The model is crucially based on the embedding of unitary $SU(2)$ representations into unitary $\SLTC$ representations. In retrospect, the use of unitary $\SLTC$ reps seems unavoidable: just as the quantum mechanics of particles naturally leads to unitary representations of Poincar\'e, so should the quantum mechanics of spacetime elements naturally lead to unitary representations of Lorentz. The unitary irreps of $\SLTC$ in the principal series are labeled by $(p,j)$, where $p$ is a real number, and $j$ is a spin (a non-negative integer or half-integer). The evidence strongly suggests that the representations which should be used are of the form $(\gamma j, j)$, where the Immirzi parameter $\gamma$ is an arbitrary real constant. Heuristically, $\gamma$ can be said to arise from an axial term which can be added to the Einstein-Hilbert action without changing the classical field equations:
\begin{align}
 S_{Holst}[e,\omega] = \int \left((e\wedge e)^* + \gamma^{-1}(e\wedge e)\right) \wedge F[\omega] \ . \label{eq:Holst}
\end{align}
$\gamma$ is conventionally taken to be positive. Nevertheless, negative values are just as legitimate, and we will consider them below. It appears that the only special values of $\gamma$ are $0$ and $\pm\infty$, the latter corresponding to the Barrett-Crane model \cite{BC}. Furthermore, it appears that a fully satisfactory construction is only possible for finite $\gamma$. In particular, the Barrett-Crane model does not capture the angular degrees of freedom in the graviton propagator \cite{BC_difficulties}.

We wish to study the parity transformation properties of the Lorentzian model with finite $\gamma$. Since there are no signs of trouble with $(C)PT$ invariance, in what follows we will consider $P$-invariance and $T$-invariance as equivalent. At least naively, the quantum theory with a fixed finite $\gamma$ is not $P$-invariant. This can be seen from \eqref{eq:Holst}, or from the fact that the $(p,j)$ representation with $p \neq 0$ is chiral: it relates e.g. $tx$ boosts with $yz$ rotations in a way which requires the right-hand rule. From the point of view of the action \eqref{eq:Holst}, the chirality of the model is a quantum effect. One can therefore hope that in the large-distance limit corresponding to semiclassical GR, parity will be restored. This argument, however, is far from robust: $\gamma$ enters crucially into the construction of spacetime itself, rather than just providing ``quantum corrections'' to its dynamics. 

Another naive expectation from the action \eqref{eq:Holst} is that a parity transformation should be equivalent to the replacement $\gamma\rightarrow -\gamma$. In the quantum theory, this statement must be handled with care. $\gamma$, and with it potential parity violation, enters the theory in two separate places. First, it enters in the vertex amplitude $A_v(h_l)$ associated with the $SU(2)$ elements $h_l$ on the links of a boundary graph. Second, it enters through the canonical commutation relations into the interpretation of boundary states $\psi(h_l)$ as 3d geometries. As we will see, it appears that the second role is the one leading to problems with parity.

A good acid test for the $P$-invariance of the theory at large distances is the calculation of the graviton propagator \cite{PropagatorConcept,PropagatorLongitudinal,EuclideanPropagator,LorentzianPropagator}. The current spinfoam model was motivated in part by the desire to get a spin-2 graviton \cite{PropagatorAsymptotics}. This seems to have been achieved \cite{PropagatorNewVertex,EuclideanPropagator,LorentzianPropagator}, and it is now time to worry about the graviton's chirality. A difference between left-handed and right-handed gravitons, like any deviation from GR at large distances, would be highly problematic for two reasons. First, we of course wish to reproduce classical GR with its empirical success. Second, on the interacting level, GR is the only unitary and Poincar\'e-invariant theory for low-energy elementary spin-2 particles. Therefore, once gravitons are present, the agreement of their behavior with GR expectations is a necessary condition for the very existence of a consistent flat-space limit. Note that neither of these issues can be settled at the propagator level: the propagator can always be made non-chiral by rescaling the graviton field components. However, this would constitute a modification to the geometric interpretation of boundary states.

Some information on parity issues is already available for the Euclidean version of the theory. First, the Euclidean EPRL vertex is manifestly invariant under the replacement $\gamma\rightarrow -\gamma$. To see this, consider the vertex amplitude for a semicoherent state:
\begin{align}
 A_{Eucl.}(j_l, \vec n_l, \vec n'_l) = \int dg^\pm \prod_l \prod_{i = \pm} \left<\vec n_l| g^i_{s(l)} (g^i_{t(l)})^{-1} |\vec n'_l\right>^{2j_l^i} \ ,
\end{align}
where $j_l^\pm = (|1 \pm \gamma|/2)j_l$. Setting $\gamma\rightarrow -\gamma$ interchanges the $j_l^+$ and $j_l^-$ factors, without changing the result. Also, the large-spin 4-simplex transition amplitude for a semicoherent boundary state \cite{EuclideanAmplitude} is composed of two parity-related critical points, with equal weight to each (plus two unwanted terms which are peculiar to the Euclidean). 

On the other hand, the Euclidean graviton propagator was calculated in the large-spin 4-simplex limit \cite{EuclideanPropagator}, and found to violate parity: it is not invariant under odd permutations of the 5 nodes, which correspond to parity-odd 4d isometries. As could be expected, invariance is restored in the $\gamma = 0$ limit. The invariance of the vertex under $\gamma\rightarrow -\gamma$ implies that the problem arises from the geometric interpretation of the boundary state. It is conceivable that this chirality is an artifact of the 4-simplex approximation, and will become negligible when larger graphs and spinfoams are taken into account. Still, the evidence is troubling, and it would be desirable to resolve the problem already at the 4-simplex level.

In parity-related questions, extrapolation from the Euclidean theory to the Lorentzian is clearly problematic. We must address the issues in the Lorentzian case directly. Important analytical progress has already been made in this direction. The semicoherent transition amplitudes in the large-spin 4-simplex limit have been studied in \cite{LorentzianAmplitude}. Using the framework developed there, the graviton propagator in the same limit was recently analyzed \cite{LorentzianPropagator}. Unfortunately, with current analytical methods the Lorentzian theory is less transparent than the Euclidean. As a result, parity invariance has not yet been successfully addressed. This is the task of the present paper. 

In a nutshell, the parity features we find for the Lorentzian theory are analogous to those found in the Euclidean. In particular, we calculate the ``semicoherent'' piece of the graviton propagator, i.e. the piece that arises from variations of the rotation and spinor variables at fixed spins. This is the part of the propagator which is potentially problematic under parity. We calculate this piece for two parity-related components of the propagator, and find that the values differ by the same ratio $e^{2\pi i/3}$ as in the Euclidean \cite{EuclideanPropagator}. Thus, the chirality problem with the graviton propagator persists in the Lorentzian model.

Our study of the Lorentzian 4-simplex was done numerically, using a Python script. The script is based on the definitions in \cite{LorentzianAmplitude}, together with the formulas for the metric insertions from \cite{LorentzianPropagator} (there is a slight discrepancy in conventions between the two papers, and we stuck to the conventions of \cite{LorentzianAmplitude}). Some technical comments regarding the script are given in the Appendix. The script files themselves are included in the arXiv submission.

In section \ref{sec:symmetry}, we demonstrate analytically that the Lorentzian vertex, like the Euclidean, is invariant under $\gamma\rightarrow -\gamma$. In the process, we find a phase discrepancy between the vertex as defined in terms of group characters (e.g. \cite{Zakopane}) and the one used in \cite{LorentzianAmplitude}. We argue that the version in \cite{Zakopane} is the more appropriate one. In section \ref{sec:numeric}, we present numerical results for the 4-simplex. These include the chirality of the graviton propagator, as well as some symmetry relations between the parity-related critical points of the transition amplitude. These symmetry relations have a very simple form, which can certainly be derived analytically. In section \ref{sec:numeric:orientation}, we present a minor correction to the geometric construction of semiclassical boundary states in \cite{LorentzianAmplitude}. In section \ref{sec:discuss}, we discuss future prospects.
 
\section{The $\gamma\rightarrow -\gamma$ symmetry of the vertex} \label{sec:symmetry}

\subsection{Derivation from the group-character definition of the vertex} \label{sec:symmetry:character}

The vertex amplitude $A_v$ as a function of $SU(2)$ elements $h_l$ on its surrounding graph is given by \cite{Zakopane}:
\begin{align}
 A_{v,\gamma}(h_l) = \int_{\SLTC^{N-1}}\prod_n dg_n \prod_l \sum_j(2j + 1)^2 \int_{SU(2)} dk_l\, \chi^j(h_l k_l)\, \chi^{\gamma j, j}(k_l g_{s(l)} g^{-1}_{t(l)}) \ , \label{eq:vertex}
\end{align}
where we made the dependence on $\gamma$ explicit in the notation. $n$ labels the graph's nodes, $l$ labels the links, and $s(l),t(l)$ are the source and target nodes of the link $l$. As usual, for a graph with $N$ nodes, we integrate over just $N-1$ Lorentz group elements $g_n$, with the remaining one fixed to the identity in order to make the integral finite. In this subsection, we will show that $A_{v,\gamma} = A_{v,-\gamma}$. For this purpose, it's convenient to think of the integration variables in \eqref{eq:vertex} as actual $\SLTC$ (or $SU(2)$) matrices, rather than abstract group elements. To streamline notations, we introduce a symbol for the $J$-conjugate of an $\SLTC$ matrix:
\begin{align}
 g^J \equiv JgJ^{-1} = (g^{-1})^\dagger \ , \label{eq:g_J}
\end{align}
where the $J$ in the second expression is the standard parity operation on 2-spinors. The operation \eqref{eq:g_J} preserves the multiplication order and the Haar measure, and leaves $SU(2)$ matrices unchanged. It can be interpreted as either a $P$ or a $T$ reflection of the Lorentz rotations.

The inequivalent $\SLTC$ matrices are fully characterized by their complex trace. In its eigenframe, such a matrix is seen to consist of a spatial rotation and a boost in the perpendicular plane. The trace then encodes the rotation angle and the boost parameter. The complex conjugate of a given trace encodes (for instance) the same rotation with the opposite boost. Intuitively, this corresponds to a symmetry of the unitary Lorentz reps:
\begin{align}
 \tr{B} = \overline{\tr{A}} \quad\Rightarrow\quad \chi^{p, j}(B) = \chi^{-p, j}(A) \ . \label{eq:symm_general}
\end{align}
We will use below a special case of this relation:
\begin{align}
 \chi^{p, j}(g^J) = \chi^{-p, j}(g) \ . \label{eq:symm_J}
\end{align}
Let us present a short proof of \eqref{eq:symm_J}. The equivalence of matrices with equal traces will then imply the general relation \eqref{eq:symm_general}. The $(p,j)$ representation of $\SLTC$ is given by functions of a 2-spinor $z$ with the homogeneity property:
\begin{align}
 f(\lambda z) = \lambda^{-1+j+ip} \bar\lambda^{-1-j+ip} f(z) \ , \label{eq:homogeneity}
\end{align}
with the group action and the Hilbert product defined by:
\begin{align}
 U_g[f](z) &= f(g^T z) \label{eq:U} \\
 \left<f_1, f_2\right> &= \int_{\mathbb{CP}^1} \Omega\, \bar f_1(z) f_2(z) \ . \label{eq:Hermitian_product}
\end{align}
The integral is over non-collinear spinors $z$, and the two-form measure $\Omega$ is defined as:
\begin{align}
 \Omega = \frac{i}{2}(z_0 dz_1 - z_1 dz_0)\wedge (\bar z_0 d\bar z_1 - \bar z_1 d\bar z_0) \ , \label{eq:Omega}
\end{align}
where $z_0$ and $z_1$ are the two components of $z$. Note that $\Omega$ is real. Now, consider the representation obtained by acting with the matrices $g^J$ instead of $g$:
\begin{align}
 \tilde U_g[f](z) = U_{g^J}[f](z) = f((g^T)^J z) \ . \label{eq:U_tilde}
\end{align}
We claim that this new representation is in fact the $(-p,j)$ representation under a change of variables. To see this, define a new set of homogeneous functions $\tilde f(z) \equiv f(Jz)$.
The expression \eqref{eq:Hermitian_product} for the Hermitian product is invariant under $z\rightarrow Jz$ (the 2-form \eqref{eq:Omega} picks up a minus sign, but this is canceled as usual by a reversal of the integration interval). In terms of $Jz$, the transformation law \eqref{eq:U_tilde} takes the form:
\begin{align}
 \tilde U_g[f](Jz) = f((g^T)^J Jz) = f(J(g^T z)) \ .
\end{align}
Therefore, the functions $\tilde f(z)$ transform under the canonical rule \eqref{eq:U}. Finally, the homogeneity of $\tilde f(z)$ is obtained from \eqref{eq:homogeneity} by interchanging $\lambda$ and $\bar\lambda$. The result is the homogeneity rule of the $(p,-j)$ representation. We conclude that the action of $g^J$ on the $(p,j)$ representation elements $f$ is isomorphic to the action of $g$ on the $(p,-j)$ representation elements $\tilde f$. As is well known, the $(p,-j)$ representation is isomorphic to the $(-p,j)$ representation, so eq. \eqref{eq:symm_J} follows.

Coming back to the vertex amplitude \eqref{eq:vertex}, let us perform a change of integration variables at the nodes:
\begin{align}
 g_n \rightarrow g_n^J \ . \label{eq:inverse_dagger}
\end{align}
This does not affect the Haar measure, or the fixing of one of the $g_n$ to the identity. Therefore, $dg_n \rightarrow dg_n$. Since $k_l = k_l^J$, the argument of the $\SLTC$ character in \eqref{eq:vertex} transforms as:
\begin{align}
 k_l g_{s(l)} g^{-1}_{t(l)}\quad \rightarrow\quad k_l g^J_{s(l)} (g^{-1}_{t(l)})^J = \left(k_l g_{s(l)} g^{-1}_{t(l)}\right)^J \ . \label{eq:kgg}
\end{align}
Using \eqref{eq:symm_J}, this implies:
\begin{align}
 \chi^{\gamma j, j}(k_l g_{s(l)} g^{-1}_{t(l)}) \rightarrow \chi^{-\gamma j, j}(k_l g_{s(l)} g^{-1}_{t(l)}) \ .
\end{align}
Since the entire operation was just a change of integration variables, we obtain the result:
\begin{align}
 A_{v,\gamma}(h_l) = A_{v,-\gamma}(h_l) \ . \label{eq:symm_SU2_vertex}
\end{align}

\subsubsection{Alternative derivation without infinite-dimensional characters}

The above derivation of the symmetry \eqref{eq:symm_SU2_vertex} made free use of the infinite-dimensional $\SLTC$ characters $\chi^{\gamma j,j}$. These are defined only in a distributional sense. Therefore, it's worth presenting an equivalent derivation which doesn't invoke these quantities. Let us consider the alternative definition of \eqref{eq:vertex} in terms of representation matrix elements \cite{Zakopane}:
\begin{align}
 A_{v,\gamma}(h_l) = \int_{\SLTC^{N-1}}\prod_n dg_n \prod_l \sum_j(2j + 1)\sum_{m,m'=-j}^j \overline{D^j(h_l)^m{}_{m'}} D^{\gamma j, j}(g_{s(l)} g^{-1}_{t(l)})^{jm'}{}_{jm} \ , \label{eq:vertex_D}
\end{align}
where $D^j(h)$ is an $SU(2)$ representation matrix in the magnetic-number basis, and $D^{\gamma j,j}(g)$ is an $\SLTC$ representation matrix in the (spin, magnetic number) basis. Once again, we can perform the substitution $g_n \rightarrow g_n^J$ on the integration variables without changing the result. This substitution sends $g_{s(l)} g^{-1}_{t(l)} \rightarrow (g_{s(l)} g^{-1}_{t(l)})^J$. Then to demonstrate the symmetry \eqref{eq:symm_SU2_vertex}, it suffices to establish the relation: 
\begin{align}
 D^{-\gamma j, j}(g)^{jm'}{}_{jm} = D^{\gamma j, j}(g^J)^{jm'}{}_{jm} \ , \label{eq:D_symmetry}
\end{align}
for an arbitrary $\SLTC$ element $g_{s(l)} g^{-1}_{t(l)} \equiv g$. To prove this relation, we will use the decomposition:
\begin{align}
 g = u_1 b(\varepsilon) u_2 \ ,
\end{align}
where $u_1,u_2 \in SU(2)$, and $b(\varepsilon)$ is a pure boost in the $tz$ plane with boost parameter $\varepsilon$:
\begin{align}
 b(\varepsilon) = \left(
  \begin{array}{ll}
    \varepsilon & 0 \\
    0 & \varepsilon^{-1}
  \end{array}\right) \ .
\end{align}
The $SU(2)$ elements $u_1,u_2$ are invariant under the $J$-conjugation. Also, the corresponding representation matrices are independent on $\gamma$, and are simply given by the spin-$j$ $SU(2)$ matrices:
\begin{align}
 D^{\gamma j, j}(u_i)^{jm'}{}_{jm} = D^{-\gamma j, j}(u_i)^{jm'}{}_{jm} = D^j(u_i)^{m'}{}_m \ .
\end{align}
It therefore suffices to demonstrate the relation \eqref{eq:D_symmetry} for the pure boost $b(\varepsilon)$. The effect of $J$-conjugation on $b(\varepsilon)$ is to send the boost parameter to its inverse: $\varepsilon\rightarrow\varepsilon^{-1}$. We must therefore show that:
\begin{align}
 D^{-\gamma j, j}\left(b(\varepsilon)\right)^{jm'}{}_{jm} = D^{\gamma j, j}\left(b(\varepsilon^{-1})\right)^{jm'}{}_{jm} \ , \label{eq:b_symmetry}
\end{align}
The matrix elements of a pure boost $b(\varepsilon)$ in the $(p,j)$ representation are explicitly known \cite{LorentzReps}. The elements of interest to us are obtained by setting $j = j' = \nu_0$ in eq. (4.11) of \cite{LorentzReps}. We get:
\begin{align}
 \begin{split}
   D^{\gamma j, j}\left(b(\varepsilon)\right)^{jm'}{}_{jm} ={}& \delta^{m'}_m \varepsilon^{2\left(1 + m + j(1 + i\gamma/2)\right)} \\
    & \cdot F\left(1 + j\left(1 + \frac{i\gamma}{2}\right),\, j + m + 1;\, 2(j + 1);\, 1 - \varepsilon^4 \right) \ , \label{eq:D_explicit}
 \end{split}
\end{align}
where $F(\alpha,\beta;x;y)$ is the hypergeometric function. The desired relation \eqref{eq:b_symmetry} can be derived directly from \eqref{eq:D_explicit}, using the following property of $F(\alpha,\beta;x;y)$:
\begin{align}
 F(\alpha,\beta;x;y) = (1 - y)^{-\beta} F\left(x - \alpha,\, \beta;\, x;\, \frac{y}{y - 1} \right) \ .
\end{align}
Reeling back the string of logic, we have thus established the symmetry \eqref{eq:symm_SU2_vertex} of the vertex amplitude.

\subsection{Derivation from semicoherent states} \label{sec:symmetry:semicoherent}

In \cite{LorentzianAmplitude}, the vertex amplitude for a semicoherent state with spins $j_{ab}$ and spinor parameters $\xi_{ab}$ is written as:
\begin{align}
  \tilde A_v(j_{ab}, \xi_{ab}, \xi_{ba}) &= (-1)^\chi \int_{\SLTC^{N-1}}\prod_n dg_n \prod_{\substack{a=s(l)\\b=t(l)}} \tilde P_{ab} \ , \label{eq:tilde_A}
\end{align}
where $n$ labels the graph's nodes, $l$ labels the links, and $a,b$ label the source and target nodes of each link. $(-1)^\chi$ is a global sign factor from combinatorics, and the ``link propagator'' $\tilde P_{ab}$ is given by:
\begin{align}
 \begin{split}
   \tilde P_{ab} &= \beta\left(\bar g_a\mathcal{I}\phi_{ab}, \bar g_b\mathcal{I}\phi_{ba}\right) \\
     &= \frac{c_{ab}}{\pi}(2j_{ab} + 1)\int_{\mathbb{CP}^1}\frac{\Omega}{\|Z_{ab}\|^2 \|Z_{ba}\|^2}
      \left(\frac{\|Z_{ba}\|}{\|Z_{ab}\|}\right)^{2ip_{ab}}\left(\frac{\left<Z_{ab}, \xi_{ab}\right>\left<-JZ_{ba}, \xi_{ba}\right>}{\|Z_{ab}\| \|Z_{ba}\|} \right)^{2j_{ab}} \ .
 \end{split} \label{eq:tilde_P_ab}
\end{align}
We refer the reader to \cite{LorentzianAmplitude} for a full explanation of this expression. The integral is performed over 2-spinors $z_{ab}$ with the measure $\Omega$ from \eqref{eq:Omega}. The $Z$ spinors are a shorthand notation for $Z_{ab} = g_a^\dagger z_{ab}$ and $Z_{ba} = g_b^\dagger z_{ab}$. The angle brackets stand for the Hermitian inner product in $\mathbb{C}^2$, and the norms $\|Z\|$ are defined with respect to that inner product. The phase factor $c_{ab}$ is given by:
\begin{align}
 c_{ab} = \frac{j_{ab} + ip_{ab}}{\sqrt{j_{ab}^2 + p_{ab}^2}} \ .
\end{align}
We set $p_{ab} = \gamma j_{ab}$, which makes all the $c_{ab}$'s equal:
\begin{align}
 c_{ab} = c_\gamma = \frac{1 + i\gamma}{\sqrt{1 + \gamma^2}} \ .
\end{align}
Note the asymmetric treatment in \eqref{eq:tilde_P_ab} of the link's source and target nodes $a$ and $b$. We may interchange them by using the symmetry property of the $\beta$ form \cite{LorentzianAmplitude}:
\begin{align}
 \beta(\phi_1, \phi_2) = (-1)^{2j}\beta(\phi_2, \phi_1) \ .
\end{align}
We then have:
\begin{align}
 \tilde A_v &= (-1)^\chi \int_{\SLTC^{N-1}}\prod_n dg_n \prod_{\substack{a=s(l)\\b=t(l)}} (-1)^{2j_{ab}} P_{ba} \label{eq:f_ba} \\
 \begin{split}
   \tilde P_{ba} &= \beta\left(\bar g_b\mathcal{I}\phi_{ba}, \bar g_a\mathcal{I}\phi_{ab} \right) \\
     &= \frac{c_{ab}}{\pi}(2j_{ab} + 1)\int_{\mathbb{CP}^1}\frac{\Omega}{\|Z_{ab}\|^2 \|Z_{ba}\|^2}
      \left(\frac{\|Z_{ab}\|}{\|Z_{ba}\|}\right)^{2ip_{ab}}\left(\frac{\left<-JZ_{ab}, \xi_{ab}\right>\left<Z_{ba}, \xi_{ba}\right>}{\|Z_{ab}\| \|Z_{ba}\|} \right)^{2j_{ab}} \ .
 \end{split} \label{eq:P_ba}
\end{align}
Now consider the change of integration variables:
\begin{align}
 g_n \rightarrow g_n^J;\quad z_{ab} \rightarrow Jz_{ab} \ .
\end{align}
The $dg_n$ and $\Omega$ integrals are invariant under these changes; so is the fixing to the identity of one of the $g_n$. The induced change in $Z_{ab}$ and $Z_{ba}$ is:
\begin{align}
 Z_{ab} \rightarrow JZ_{ab};\quad Z_{ba} \rightarrow JZ_{ba} \ .
\end{align}
This turns \eqref{eq:P_ba} into:
\begin{align}
  \tilde P_{ba} \rightarrow \frac{c_{ab}}{\pi}(2j_{ab} + 1)\int_{\mathbb{CP}^1}\frac{\Omega}{\|Z_{ab}\|^2 \|Z_{ba}\|^2}
     \left(\frac{\|Z_{ab}\|}{\|Z_{ba}\|}\right)^{2ip_{ab}}\left(\frac{\left<Z_{ab}, \xi_{ab}\right>\left<JZ_{ba}, \xi_{ba}\right>}{\|Z_{ab}\| \|Z_{ba}\|} \right)^{2j_{ab}} \ .
\end{align}
This is the same as $P_{ab}$, except for the sign on $JZ_{ba}$ and an inversion of the $(\|Z_{ba}\|/\|Z_{ab}\|)^{2ip_{ab}}$ factor. This last difference corresponds to flipping the sign of $p_{ab}$, or equivalently of $\gamma$. Keeping in mind also the dependence of $c_{ab}$ on $p_{ab}$, we get:
\begin{align}
 \tilde P_{ba}(\gamma) = (-1)^{2j_{ab}}c^2_{ab} \tilde P_{ab}(-\gamma) \ .
\end{align}
The sign factor cancels nicely with the one in \eqref{eq:f_ba}, giving:
\begin{align}
 \tilde A_{v,\gamma}(j_{ab}, \xi_{ab}, \xi_{ba}) = c^{2L}_\gamma \tilde A_{v,-\gamma}(j_{ab}, \xi_{ab}, \xi_{ba}) \ , \label{eq:symm_barrett_vertex}
\end{align}
where $L$ is the number of links in the graph. There is a discrepancy between the symmetries \eqref{eq:symm_barrett_vertex} and \eqref{eq:symm_SU2_vertex}. It suggests that the vertex amplitudes \eqref{eq:vertex} and \eqref{eq:tilde_A} are not quite the same. Since \eqref{eq:symm_SU2_vertex} is a cleaner symmetry, it appears sensible to adopt a version of the semicoherent amplitude \eqref{eq:tilde_A} that respects it. This amounts to omitting the $c_{ab}$ factors in \eqref{eq:tilde_A}-\eqref{eq:tilde_P_ab} and defining:
\begin{align}
 \begin{split}
   & A_v(j_{ab}, \xi_{ab}, \xi_{ba}) = (-1)^\chi \int_{\SLTC^{N-1}}\prod_n dg_n \prod_{\substack{a=s(l)\\b=t(l)}} P_{ab} \\
   & P_{ab} = \frac{1}{\pi}(2j_{ab} + 1)\int_{\mathbb{CP}^1}\frac{\Omega}{\|Z_{ab}\|^2 \|Z_{ba}\|^2}
      \left(\frac{\|Z_{ba}\|}{\|Z_{ab}\|}\right)^{2ip_{ab}}\left(\frac{\left<Z_{ab}, \xi_{ab}\right>\left<-JZ_{ba}, \xi_{ba}\right>}{\|Z_{ab}\| \|Z_{ba}\|} \right)^{2j_{ab}} \ .
 \end{split} \label{eq:A_semicoherent}
\end{align}
This is the definition that we'll adopt in the numerical analysis below. As we'll see in section \ref{sec:numeric:crit_points}, the omission of the $c_{ab}$ factors also leads to a more symmetric relation between parity-related amplitudes at fixed $\gamma$.

\subsection{Flipping $\gamma$ at each link separately} \label{sec:symmetry:per_link}

The work reported in this section began with the hope that the vertex is \emph{not} invariant under $\gamma\rightarrow -\gamma$. Then a modified vertex of the form $A_{v,\gamma} + A_{v,-\gamma}$ could have resolved the chirality problems. As we've seen, this is not the case, and the proposed modification is trivial. 

One may consider an alternative modification, which we now briefly discuss. $\gamma$ is traditionally taken to be a global parameter. However, its actual usage in the vertex amplitude \eqref{eq:vertex} or \eqref{eq:A_semicoherent} is to raise the representations on each link from $SU(2)$ to $\SLTC$ - an operation which can be considered for each link separately. In particular, we can imagine using different signs of $\gamma$ on different links. To restore the symmetry between the links, we must then sum over all $2^L$ such possibilities.

The arguments of subsections \ref{sec:symmetry:character}-\ref{sec:symmetry:semicoherent}, which demonstrated invariance under a \emph{global} flip of $\gamma$, no longer apply. Therefore, this ``per-link'' flipping of $\gamma$ may be a genuine and interesting modification of the model. However, we must note that in the large-spin limit studied in \cite{LorentzianAmplitude}, it degenerates back to the ``global'' flip $A_{v,\gamma} + A_{v,-\gamma}$. Indeed, it was shown in \cite{LorentzianAmplitude} that to avoid exponential suppression, $p_{ab}$ must be proportional to $j_{ab}$ with the same coefficient on all the links. Therefore, mixed terms with $\gamma$ on some links and $-\gamma$ on the others won't contribute to the asymptotic amplitude.

\section{Numerical results in the large-spin 4-simplex limit} \label{sec:numeric}

\subsection{Introduction to the 4-simplex geometry} \label{sec:numeric:intro}

We now turn to the parity structure of the large-spin 4-simplex amplitudes. We wrote a numerical script for the task. The script works with the semicoherent states described in \cite{LorentzianAmplitude}, but without the $c_{ab}$ phase factors as we discussed in section \ref{sec:symmetry:semicoherent}. The script is used to generate appropriate boundary-state parameters $(j_{ab}, \xi_{ab})$ from simpler geometric data, to find the two critical points $(g_a, z_{ab})_\pm$ for the transition amplitude, to calculate the Hessian of the ``action'' at these points, and finally to calculate the ``semicoherent'' piece of the graviton propagator.

In the Euclidean calculation of the graviton propagator \cite{EuclideanPropagator}, the authors used a regular 4-simplex for maximal symmetry. In a Lorentzian signature, a regular 4-simplex doesn't exist. We therefore work with the next most symmetric possibility, i.e. an isosceles 4-simplex. It's composed of a regular ``base'' tetrahedron in the $t = 0$ hyperplane centered at the origin, and 4 isosceles ``side'' tetrahedra whose apexes meet at a point $(h,0,0,0)$. In turn, each of the isosceles tetrahedra is composed of a regular base triangle (where they touch the base tetrahedron) and 3 isosceles side triangles (where they touch each other). The base tetrahedron is labeled as node no. $0$, and the side tetrahedra are labeled as nodes $1\dots 4$.

\begin{figure}%
\centering%
\includegraphics[scale=0.5]{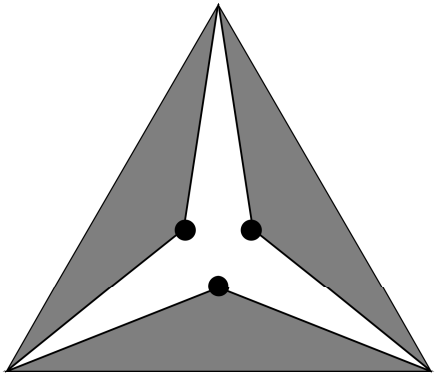} \\
\caption{A dimensionally-reduced depiction of the tetrahedra (here, triangles) that define the 4-simplex boundary state. The four (here, three) isosceles side tetrahedra are attached to the regular base tetrahedron from the inside. When the side tetrahedra are boosted by appropriate angles, the circled vertices join together. These boosts correspond to the $\SLTC$ matrices $g_a$ at the two critical points.}
\label{fig:simplex} 
\end{figure}

Figure \ref{fig:simplex} depicts (in one fewer dimension) the location of the tetrahedra in the ``source'' 3d space, where the boundary state is defined. Here again, the base tetrahedron is centered at the origin. The side tetrahedra have a smaller height, and are glued to the base tetrahedron's faces from the \emph{inside} (see section \ref{sec:numeric:orientation} on this point). The $\SLTC$ rotation matrices $g_a$ at the two critical points leave the base tetrahedron intact, while the side tetrahedra are boosted around their base triangles until their side triangles and apexes meet. The difference between the two critical points is that one boosts the side tetrahedra into positive $t$, and the other into negative $t$.

We note that exchanging two tetrahedra among the equivalent set $1\dots 4$ amounts to a parity-odd isometry on the 4-simplex. For instance, we can (and do) choose 4d axes so that tetrahedra $0,3,4$ are centered on the $z = 0$ hyperplane, while tetrahedra $1,2$ are centered at opposite values of $z$. Then the exchange $1\leftrightarrow 2$ corresponds to the parity-odd isometry $z \rightarrow -z$. This correspondence between parity and the exchange of two tetrahedra will play a crucial role in the analysis of the graviton propagator in section \ref{sec:numeric:propagator}.

\subsection{Description of the numerical calculation} \label{sec:numeric:description}

In this subsection, we describe the logical flow of the numerical calculation. Several details more related to programming than to geometry are deferred to the Appendix.

The recipe for semiclassical amplitudes in \cite{LorentzianAmplitude} begins with the parameters $(j_{ab}, \xi_{ab})$ of the boundary state. These are then translated into a discrete 4-simplex geometry by the critical points of the integrals \eqref{eq:A_semicoherent}, if the boundary parameters admit such critical points. The geometry at the critical points may be Euclidean, Lorentzian or degenerate (i.e. effectively 3-dimensional). In the present context, our goal is slightly different. We wish to start with a boundary state which corresponds to a Lorentzian 4-simplex, and use this as a background for the graviton propagator. Therefore, the calculation works ``backwards'': we start by specifying the desired geometry of the Lorentzian 4-simplex at the critical points, and calculate from that the required parameters $(j_{ab}, \xi_{ab})$ of the boundary state.

The vertex positions of the base tetrahedron are hard-coded into the script. Its edge length is normalized to 1. This is the only hard-coded geometric data. All the rest is calculated dynamically in order to minimize human error, in particular with regard to signs and orientations. The script accepts as an input the rapidity with which the side tetrahedra are to be boosted at the critical points. From this we calculate their height and locations in the 3d source space, i.e. their vertex coordinates prior to boosting. 

Once the coordinates of the tetrahedra are known, the parameters $(j_{ab}, \xi_{ab})$ of the boundary state are calculated, following the prescription of \cite{LorentzianAmplitude} in reverse (up to the caveat in section \ref{sec:numeric:orientation}). We take the spins $j_{ab}$ to simply equal the corresponding triangle areas. The normalization is irrelevant for our purposes, and no harm is done by the non-half-integer values - in the stationary-phase limit, only the relative sizes of the spins matter. For $a < b$, we choose arbitrarily the phase of $\xi_{ab}$ according to the convention $(e^{-i\varphi/2}\cos(\theta/2), e^{i\varphi/2}\sin(\theta/2))$. The phase of $\xi_{ba}$ is then determined geometrically by the prescription in \cite{LorentzianAmplitude}.

We calculate the rotation matrices $g^\pm_a$ at the critical points from the geometry of the 4-simplex. The critical values $z^\pm_{ab}$ of the $z_{ab}$ spinors are then determined from the first critical-point equation in \cite{LorentzianAmplitude}, with the second component of each spinor normalized to 1.

The major task, and the one which drove us to a numerical treatment in the first place, is to calculate the Hessian $H$ of the ``action'' $S$ \cite{LorentzianAmplitude} at the two critical points. For that purpose, we must first encode the integration variables $(g_a, z_{ab})$ into a set of $24+20 = 44$ non-redundant real quantities. Also, it's useful to choose these quantities so that the Haar and $\Omega$ measures become trivial around the critical points. We do this as follows. For the rotation matrices, we first note that only 4 of them should be integrated over. We therefore keep $g_0 = 1_{2\times 2}$, and encode only the four others. For each of those, we encode not $g_a$ itself, but the matrix $\tilde g_a = g_a(g^\pm_a)^{-1}$, which represents the deviation from the critical point. As the 6 independent real components of $\tilde g_a$, we use the real and imaginary parts of $((\tilde g_a)_{01} + (\tilde g_a)_{10})/2$, $i((\tilde g_a)_{01} - (\tilde g_a)_{10})/2$ and $(\tilde g_a)_{00}$. Near the identity, these correspond to the Cartesian components of the rotation and boost generators, which ensures even spacing under the Haar measure. To encode the $z_{ab}$ spinors, we simply use the real and imaginary parts of $(z_{ab})_0$, after normalizing $(z_{ab})_1$ to 1.

With these ingredients in place, we calculate the Hessian by straightforward numerical differentiation. We will use the determinant of $H$ for the relative weights of the critical points in the amplitude, and its inverse for the graviton propagator.

We next calculate the gradients of the metric insertion functions $q_n^{ab}$ for $a,b \neq n$ at the two critical points. The prescription for the metric insertions is taken from \cite{LorentzianPropagator}, with two trivial modifications: to accommodate for all the cases $a < b$, $a = b$ and $a > b$, and to correct for the $ab \leftrightarrow ba$ discrepancy in conventions between \cite{LorentzianPropagator} and \cite{LorentzianAmplitude}. This results in the following formula for $q_n^{ab}(g_a, z_{ab})$: 
\begin{align}
 q_n^{ab} = &\left\{
  \begin{array}{ll}
    (\gamma j_{na})^2 & \ a = b \\
    \vec A_{na}\cdot \vec A_{nb} & \ a \neq b                 
  \end{array} \right. \\
 \vec A_{na} = \gamma j_{na} \cdot &\left\{
  \begin{array}{ll}
    \displaystyle{\frac{\left<\vec\sigma Z_{na}, \xi_{na}\right>}{\left<Z_{na}, \xi_{na}\right>}} & \ n < a \\[1em]
    \displaystyle{\frac{\left<\vec\sigma(JZ_{na}), \xi_{na}\right>}{\left<JZ_{na}, \xi_{na}\right>}} & \ n > a                 
  \end{array} \right. \ ,
\end{align}
where $\vec\sigma$ are the Pauli matrices. 

At each of the two critical points, we contract the gradients of $q_n^{ab}$ with the inverse Hessian to obtain the semicoherent graviton propagator:
\begin{align}
 G_{mn}^{(ab)(cd)} = (H^{-1})^{ij} (q_m^{ab})'_i (q_n^{cd})'_j \ . \label{eq:G}
\end{align}
The propagator component $G_{mn}^{(ab)(cd)}$ represents the correlator between the metric elements $\underline{ma}\cdot\underline{mb}$ and $\underline{nc}\cdot\underline{nd}$. The indices $i,j$ in \eqref{eq:G} run over the independent components of $(g_a, z_{ab})$.

Two caveats are in order. First, the script only handles the case $m \neq n$ and $a,b,c,d \neq m,n$, when the double metric insertion is simply given by the product of two single insertions. Second, as mentioned in the Introduction, the quantity that we're computing is not the full graviton propagator, but only its ``semicoherent'' piece. The full propagator would have been obtained from a fully coherent boundary state, with a weighted sum over spins. This sum would pick just one of the two critical points, and add a contribution to \eqref{eq:G} from the derivatives with respect to $j_{ab}$. In fact, this contribution is the one containing the standard Regge propagator. We ignore it in the present context, because we are interested in parity violation, and that comes from the derivatives with respect to $(g_a, z_{ab})$. Thus, we are calculating only the potentially troublesome part of the propagator. It can be cleanly separated from the $j_{ab}$-derivatives contribution, due to the vanishing of the Hessian elements $H_{jg}, H_{jz}, H_{j\bar z}$ \cite{LorentzianPropagator}.

We stress that all of the above can be done analytically. In particular, numerical differentiation is not necessary: not only the relevant functions, but also their derivatives are worked out analytically in \cite{LorentzianPropagator}. The virtue of the numerics in this case is purely tactical - it provides a shortcut to some basic answers, allowing us to avoid tedious and error-prone derivations. Also, it will be able to provide a useful check for future analytical results. In fact, the hands-on numerics has already helped us uncover a minor problem in earlier analytical work, as we now describe.

\subsection{A note on the orientation of the side tetrahedra} \label{sec:numeric:orientation}

The critical-point equations on $(g_a, z_{ab})$ do not have solutions for generic boundary states $(j_{ab}, \xi_{ab})$. In order to correspond to a semiclassical Regge geometry, the boundary-state parameters must satisfy certain constraints. These constraints have a geometric interpretation, as described respectively in \cite{EuclideanAmplitude} and \cite{LorentzianAmplitude} for the Euclidean and Lorentzian models. In the Euclidean, the $\xi_{ab}$ should be understood as the spinor ``square roots'' of the outgoing face normals of the tetrahedra, embedded in 3d space. Then the critical-point solutions correspond to folding these tetrahedra into a 4-simplex using 4d rotations. The same procedure was then carried over to the Lorentzian model in \cite{LorentzianAmplitude}. As we noticed while looking for critical points in the numerics, there is in fact a subtle difference between the two cases which must be taken into account.

For concreteness, consider the isosceles 4-simplex described in sections \ref{sec:numeric:intro}-\ref{sec:numeric:description}. In the Euclidean model, its tetrahedra can be embedded in 3d space as a four-pointed ``star'': the base tetrahedron is situated at the center, with the side tetrahedra glued to its faces from the outside. To fold this into a 4-simplex, the side tetrahedra must be rotated along their heights into the 4th dimension by some obtuse angle. Now, in the Lorentzian case such a rotation is not continuous with the identity: it would require passing through null and timelike configurations. In other words, it isn't part of $\SLTC$. To fix this, we must glue the side tetrahedra to the \emph{inside} of the base tetrahedron, as in figure \ref{fig:simplex}. That way, they can be folded together by a finite boost.

Now, the critical-point equations in \cite{LorentzianAmplitude} require the spinors $\xi_{ab}$ and $J\xi_{ba}$ to point in the same direction after the 4d rotations. In other words, $\xi_{ab}$ and $\xi_{ba}$ should point in opposite directions. This is the case in the Euclidean setup, with all the $\xi$'s pointing along the outgoing normals of their respective tetrahedra. However, now that we took the side tetrahedra to lie inside the base tetrahedron, the base and side outgoing normals are parallel rather than antiparallel. This can be cured by taking the $\xi$'s for the side tetrahedra (or the base tetrahedron) to point along the \emph{ingoing} normals to their respective faces.

This conclusion can be generalized to arbitrary Lorentzian 4-polytopes with spacelike boundaries. The boundary polyhedra can be classified according to the time-orientation of their outgoing 4d normal (not to be confused with the face normals discussed above). In constructing the semiclassical boundary state, we must take the $\xi$ spinors for the past-pointing polyhedra (say) to point along the outgoing face normals, and for the future-pointing polyhedra - along the ingoing face normals. This was the prescription we used in the numerical script.

\subsection{The coefficients of the two critical points} \label{sec:numeric:crit_points}

In \cite{LorentzianAmplitude}, the asymptotic analysis of the Lorentzian 4-simplex amplitude (with Lorentzian critical points) ends with the following expression:
\begin{align}
 A_v = (-1)^{\chi + M}\left(N_+ e^{iS_{Regge}} + N_- e^{-iS_{Regge}}\right) \ , \label{eq:asymptotic}
\end{align}
where the sign factor comes from combinatorics, and $S_{Regge}$ is the Regge action:
\begin{align}
 S_{Regge} = \gamma\sum_{a < b} j_{ab} \Theta_{ab} \ ,
\end{align}
with $\Theta_{ab}$ the dihedral angles (actually, boost parameters). The coefficients $N_\pm$ of the two parity-related critical points are given by:
\begin{align}
 N_\pm = \left(2^{36}\pi^{12}\prod_{a < b} j_{ab}\right) \left(\frac{\Omega}{\sqrt{\det H}}\prod_{a < b}\frac{1}{\|Z_{ab}\|^2 \|Z_{ba}\|^2}\right)_\pm \ , \label{eq:N}
\end{align}
where the piece in the second parentheses is the one which depends on the critical point. We again omitted the $c_{ab}$ phase factors, as discussed in section \ref{sec:symmetry:semicoherent}. The authors of \cite{LorentzianAmplitude} left the ratio $N_+ / N_-$ as an open question. We studied this question using the numerical script. Specifically, the script calculates the quantity $\left(\sqrt{\det H}\prod \|Z_{ab}\|^2 \|Z_{ba}\|^2\right)^{-1}$ at the two critical points, keeping in mind that the $\Omega$ measure is trivial in our conventions. The numbers point clearly towards the simple result:
\begin{align}
 N_+ = \overline{N_-} \ . \label{eq:N_conjugate}
\end{align}
In other words, the amplitude \eqref{eq:asymptotic} is \emph{real}. In particular, the magnitudes $|N_+| = |N_-|$ are equal. Furthermore, at small $\gamma$ we have:
\begin{align}
 \gamma\rightarrow 0 \quad \Rightarrow \quad N_+ = \overline{N_+} = N_- = \overline{N_-} \ . \label{eq:N_real}
\end{align}
Here and below, we present extrapolations from numerical results as exact analytical relations. In all such cases, what's actually implied is that the LHS and RHS agree to an accuracy of several digits, for a representative sample of points in the parameter space.

The results \eqref{eq:N_conjugate}-\eqref{eq:N_real} may be compared with the Euclidean situation \cite{EuclideanAmplitude}, where the coefficients of $\exp(iS_{Regge})$ and $\exp(-iS_{Regge})$ are equal for arbitrary $\gamma$ (and two additional terms appear). Note that the simple relation \eqref{eq:N_conjugate} is sensitive to the overall phase of the boundary state. It holds when the overall phase of $\xi_{ab}$ and $\xi_{ba}$ is fixed by the 3d rotation which makes the corresponding triangles congruent, as prescribed in \cite{LorentzianAmplitude}. The result is also dependent on our omission of the $c_{ab}$ phase factors in \eqref{eq:A_semicoherent} and \eqref{eq:N}.

As a cross-check, the numerical results for $N_\pm$ confirm the $\gamma\rightarrow -\gamma$ symmetry claimed in section \ref{sec:symmetry}. Concretely, the identity $N_\pm(\gamma) = N_\mp(-\gamma)$ is seen to hold.

\subsection{The semicoherent graviton propagator} \label{sec:numeric:propagator}

Our main object of interest is the graviton propagator, in particular its parity-related components. More specifically, as discussed in section \ref{sec:numeric:description}, we are calculating the ``semicoherent'' propagator \eqref{eq:G}, where the summed indices $i,j$ run over the independent components of $(g_a, z_{ab})$ and \emph{not} over the spins $j_{ab}$. We calculate the quantity \eqref{eq:G} at the two critical points separately, noting that only one will remain after the sum over spins. Also, we reiterate that \eqref{eq:G} is valid only for $m \neq n$ and $a,b,c,d \neq m,n$.

As a first result of this calculation, we again see a confirmation of the $\gamma\rightarrow -\gamma$ symmetry, i.e. $G_{mn}^{(ab)(cd)}(\gamma) = G_{mn}^{(ab)(cd)}(-\gamma)$. To go further, we restrict our attention to the components $G_{0n}^{(ab)(cd)}$, with $1 \leq n,a,b,c,d \leq 4$ and $a,b,c,d \neq n$. As expected from rotational invariance, all even permutations of the equivalent nodes $(1,2,3,4)$ result in the same value for $G_{0n}^{(ab)(cd)}$. However, odd permutations, which are related to the even ones by spatial parity, result in a different value. As representatives of these two equivalence classes, we pick $G_{01}^{(23)(24)}$ and $G_{01}^{(24)(23)}$. We now evaluate these components at the \emph{same} critical point (the ``positive'' one), since that's how they will enter the full coherent propagator. In figure \ref{fig:G_of_gamma}, we plot the real and imaginary parts of these quantities (divided by $\gamma^4$), as well as their complex phase, as functions of $\gamma$. We use fixed values for the spins and a fixed boost parameter determining the shape of the 4-simplex. From the numbers and from the graphs, we make the following observations:
\begin{figure}%
\centering%
\subfigure[{The real and imaginary parts of two parity-related propagator components. We divide by $\gamma^4$, because the corresponding graphs for the Euclidean regular 4-simplex are straight lines.}]%
{\includegraphics[scale=0.75]{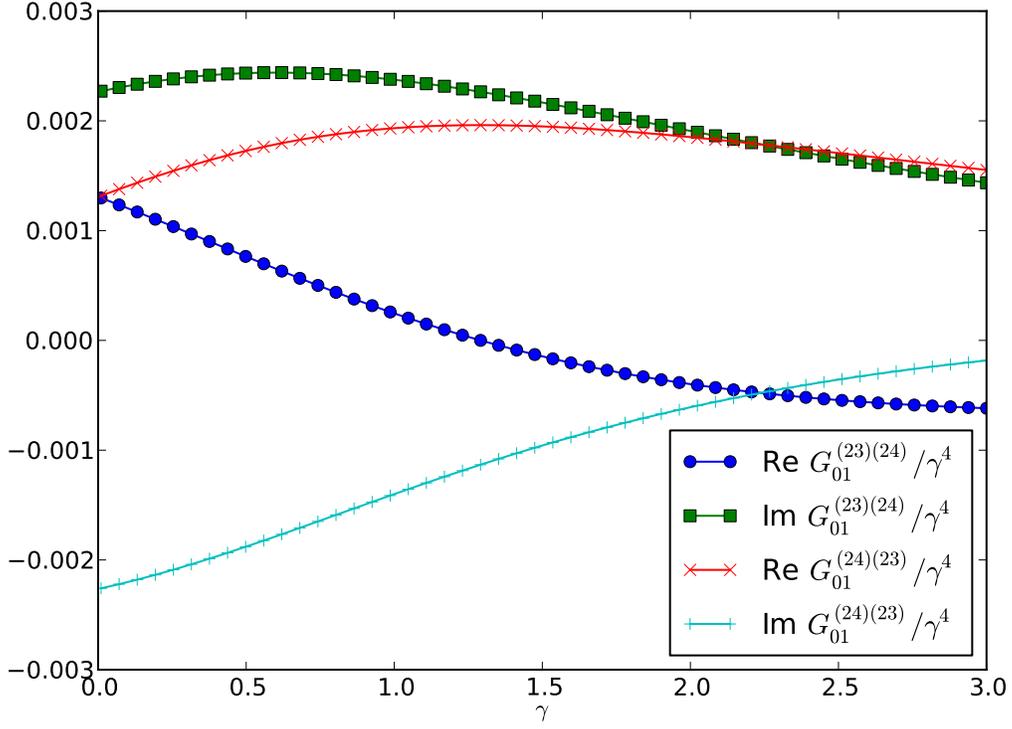} \label{fig:G_of_gamma:components}} \\
\subfigure[{The complex phase of the same two propagator components.}]%
{\includegraphics[scale=0.75]{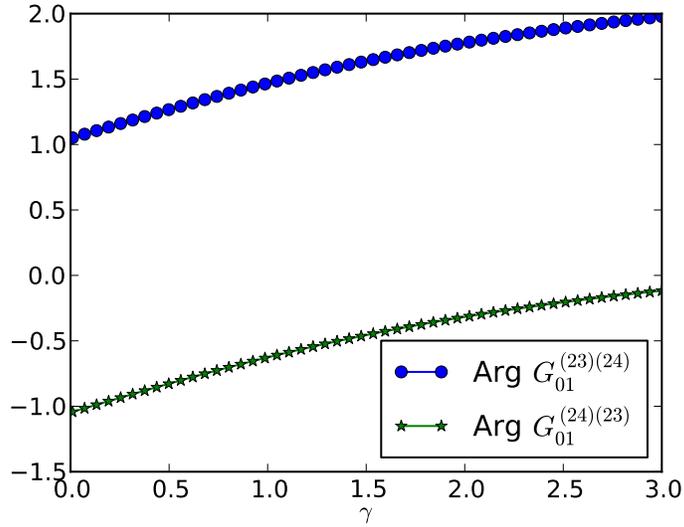}} \\
\caption{The ``semicoherent'' piece of the 4-simplex graviton propagator at one of the critical points as a function of the Immirzi parameter. The boost parameter of the side tetrahedra is fixed to $1$.}
\label{fig:G_of_gamma} 
\end{figure}%
\begin{enumerate}
 \item The ratio of the two parity-related propagator components is: 
 \begin{align}
   G_{01}^{(23)(24)}/G_{01}^{(24)(23)} = e^{2\pi i/3} \ , \label{eq:G_phase}
 \end{align}
 exactly as in the Euclidean case \cite{EuclideanPropagator}. In particular, we see that this result holds also for 4-simplices which are not regular.
 \item At small $\gamma$, the semicoherent propagator scales as $\gamma^4$. This can be seen from the fact that $G/\gamma^4$ remains regular as $\gamma \rightarrow 0$. The same scaling law has been found to hold in the Euclidean model \cite{EuclideanPropagator}. Since the $P$-invariant Regge piece of the propagator scales as $\gamma^3$ \cite{LorentzianPropagator}, it will dominate at small $\gamma$.
 \item At small $\gamma$, the two parity-related components become each other's conjugates: 
 \begin{align}
   \gamma\rightarrow 0 \quad \Rightarrow \quad G_{01}^{(23)(24)} = \overline{G_{01}^{(24)(23)}} \ . \label{eq:G_conjugate}
 \end{align}
 The same was found in \cite{EuclideanPropagator} for the Euclidean model. 
 \item In the Euclidean regular 4-simplex \cite{EuclideanPropagator}, the semicoherent piece of the propagator is proportional to $\gamma^4(7 - i\sqrt{15}\,\gamma)$. With such a dependence, the graphs in figure \ref{fig:G_of_gamma:components} would have been straight lines. We conclude that the Lorentzian propagator does not have the same polynomial dependence on $\gamma$.
\end{enumerate}
Finally, another regularity emerges from comparing the propagator components at the two critical points. Specifically, a propagator component at one point equals the complex conjugate of the parity-related component at the other point. For example:
\begin{align}
 \left(G_{01}^{(23)(24)}\right)_+ = \overline{\left(G_{01}^{(24)(23)}\right)_-} \ . \label{eq:G_two_points}
\end{align}

We stress that the above results for the graviton propagator do not depend on the choice of phases for the spinors $\xi_{ab}$ defining the boundary state. This is in contrast to the results for the transition amplitude in section \ref{sec:numeric:crit_points}. The reason for this robustness is that the propagator is a ratio \cite{EuclideanPropagator,LorentzianPropagator} between  two amplitudes - one with the metric insertions and one without. Any phase factors in the boundary state cancel when taking this ratio. Thus, it appears unlikely that the parity-violating result \eqref{eq:G_phase} is an artifact of any phase or orientation choices in the definition of the boundary state (other than the global orientation, which is the subject of interest). 

\section{Discussion} \label{sec:discuss}

We have seen that the Lorentzian 4-simplex graviton propagator is not parity-invariant, exactly as in the Euclidean case. We've also shown that flipping the sign of $\gamma$ has no effect on the spinfoam vertex formula, so it cannot help to resolve the problem. If LQG is to be taken seriously as reproducing GR at large distances, this issue must be addressed. 

Perhaps the problem will go away once larger graphs and spinfoams are taken into account. Such calculations should be performed in order to find out. Statistically, one may expect large graphs with small spins on the links to dominate over small graphs with large spins. Therefore, a good place to start may be the single-vertex approximation for a large graph with spin-$1/2$ links. Perhaps the spin-$1/2$ 4-simplex will be a good warmup exercise, though it isn't directly relevant to the large-distance limit. 

Another possibility is that the correlator $G_{mn}^{(ab)(cd)}$ between two single nodes is not the correct quantity to describe the large-distance graviton propagator. In the crude 4-simplex approximation, there's really no other choice. However, given a large boundary graph with many Planck-scale nodes, we may consider a correlator that's smeared over many nearby nodes around each of the two endpoints. In fact, one might say that calculating a large-distance propagator between points defined at a Planckian resolution is outright suspect.

We should note here that the parity problem, in both the Euclidean and Lorentzian models, goes away in the limit $\gamma\rightarrow 0$. In fact, the evidence so far seems to indicate that the large-spin, small-$\gamma$ limit precisely reproduces classical Regge gravity \cite{VanishCosmology,Vanish,LorentzianPropagator}. We are not sure what to conclude from this. One difficulty should certainly be noted: in the aforementioned limit, Regge gravity is indeed reproduced precisely, e.g. with no renormalization of Newton's constant. This stands in tension with the LQG black-hole entropy calculations \cite{BH1,BH2,BH3}, where agreement with the Bekenstein formula demands either an order-1 value for $\gamma$ or a $\gamma$-proportional renormalization ratio $G_{IR}/G_{UV}$.

It is possible that none of the above solves the issue, and that the theory must be modified. As we've seen, the vertex formula \eqref{eq:vertex} already appears $P$-invariant. Therefore, the necessary modification may lie in the geometric interpretation of the boundary states. Rescaling the graviton field components to obtain a $P$-symmetric propagator falls within this category. It is an extreme possibility, as it amounts to changing the time-honored kinematics of LQG. Finally, it may be that the problem will be solved by large graphs with small spins together with the per-link modification proposed in section \ref{sec:symmetry:per_link}.

Full analytical understanding of the 4-simplex system is an important task in itself. Our numerical observations \eqref{eq:N_conjugate}-\eqref{eq:N_real} and \eqref{eq:G_phase}-\eqref{eq:G_two_points} can and should be derived analytically. At present this seems like tedious work. Given the simplicity of the final results, we are probably missing some crucial ideas. As an intermediate step, symbolic math software may be considered.

The numerical 4-simplex script presented here can be used for questions other than parity invariance. More generally, LQG is bursting with fundamental questions for which numerical answers will be vastly better than nothing. We should look forward towards more comprehensive numerical tools, which are still largely missing.

\section*{Acknowledgements}		

I am grateful to Carlo Rovelli for discussions and a vegan pizza. The Les Houches summer school ``Theoretical physics to face the challenge of LHC'' provided a wonderful atmosphere during part of this project. The work is supported in part by the Israeli Science Foundation center of excellence, by the US-Israel Binational Science Foundation (BSF), and by the German-Israeli Foundation (GIF). The numerics were written in Python, using the NumPy and PyLab modules and the Spyder development environment. Credit for these tools goes to the open-source scientific computing community.

\appendix
\section{Technical comments regarding the numerical script} \label{app:script} 

\subsection{Methods of invoking the script}

In writing the numerical script, efficiency was consistently sacrificed for simplicity. The 4-simplex calculations, with textual output, are carried out by the script file \texttt{spinfoam\_4simplex.py}. The plots in figure \ref{fig:G_of_gamma} were produced using the script file \texttt{G\_plot.py}. The main script \texttt{spinfoam\_4simplex.py} can be run as a standalone (see the \texttt{usage\_string()} function), or from another program using the \texttt{calc\_4simplex()} function. A '\texttt{verbose}' binary flag causes the script to print out in detail the parameters of the boundary state and the critical points. A '\texttt{debug}' flag prints out some consistency checks: that the critical-point equations are satisfied, that the gradient of $S$ at the critical points is small, etc. A '\texttt{crit-points}' option can instruct the script to process just one critical point, as is relevant for the fully coherent boundary state, or none, if we are only interested in the boundary state parameters.

\subsection{Vertex placement and symmetry planes of the base tetrahedron}

In section \ref{sec:numeric:description}, we mentioned that the base tetrahedron's vertices are hard-coded so as to obtain a regular tetrahedron with edge length $1$ centered at the origin. This leaves us with the freedom to choose the orientation of the tetrahedron. Our choice was to place one of the face normals along the $x$ axis, and to align one of the symmetry planes with the $xy$ plane. In making this choice, it was important that none of the face normals are parallel to the $z$ axis - this is the only failsafe against singularities due to a vanishing component of a 2-spinor. 

\subsection{Choice of differentiation step}

In our calculation of the gradients and the Hessian at the critical points, the differentiation step is hard-coded. The optimal order of magnitude for the step was found by a trial-and-error process. There are two opposing constraints at work here. On one hand, the step should be large enough to be well above the floating-point accuracy. For this, we ensure that the results of differentiation are insensitive to order-of-magnitude changes in the step. On the other hand, the step should be small enough to avoid smearing the results. For this, we ensure the smallness of the ``action'' gradients at the critical points. We found a range of several orders of magnitude which satisfies both requirements, and chose a step near the middle of that range.

This manual choice of the step should be kept in mind (and hopefully improved upon) if one wishes to adapt the code for a different calculation.


\begin{thebibliography} {99}

\bibitem{FK}
  L.~Freidel, K.~Krasnov,
  ``A New Spin Foam Model for 4d Gravity,''
  Class.\ Quant.\ Grav.\  {\bf 25}, 125018 (2008).
  [arXiv:0708.1595 [gr-qc]].

\bibitem{EPRL}
  J.~Engle, E.~Livine, R.~Pereira and C.~Rovelli,
  ``LQG vertex with finite Immirzi parameter,''
  Nucl.\ Phys.\  B {\bf 799}, 136 (2008)
  [arXiv:0711.0146 [gr-qc]].

\bibitem{BC} 
  J.~W.~Barrett and L.~Crane,
  ``A Lorentzian signature model for quantum general relativity,''
  Class.\ Quant.\ Grav.\ \ {\bf 17}, 3101  (2000)
  [gr-qc/9904025].

\bibitem{BC_difficulties} 
  E.~Alesci and C.~Rovelli,
  ``The Complete LQG propagator. I. Difficulties with the Barrett-Crane vertex,''
  Phys.\ Rev.\ D\ {\bf 76}, 104012  (2007)
  [arXiv:0708.0883 [gr-qc]].

\bibitem{PropagatorConcept} 
  C.~Rovelli,
  ``Graviton propagator from background-independent quantum gravity,''
  Phys.\ Rev.\ Lett.\ \ {\bf 97}, 151301  (2006)
  [gr-qc/0508124].

\bibitem{PropagatorLongitudinal} 
  E.~Bianchi, L.~Modesto, C.~Rovelli and S.~Speziale,
  ``Graviton propagator in loop quantum gravity,''
  Class.\ Quant.\ Grav.\ \ {\bf 23}, 6989  (2006)
  [arXiv:gr-qc/0604044 [gr-qc]].

\bibitem{EuclideanPropagator}
  E.~Bianchi, E.~Magliaro and C.~Perini,
  ``LQG propagator from the new spin foams,''
  Nucl.\ Phys.\  B {\bf 822}, 245 (2009)
  [arXiv:0905.4082 [gr-qc]].

\bibitem{LorentzianPropagator} 
  E.~Bianchi and Y.~Ding,
  ``Lorentzian spinfoam propagator,''
  arXiv:1109.6538 [gr-qc].

\bibitem{PropagatorAsymptotics} 
  E.~Alesci and C.~Rovelli,
  ``The Complete LQG propagator. II. Asymptotic behavior of the vertex,''
  Phys.\ Rev.\ D\ {\bf 77}, 044024  (2008)
  [arXiv:0711.1284 [gr-qc]].

\bibitem{PropagatorNewVertex} 
  E.~Alesci, E.~Bianchi and C.~Rovelli,
  ``LQG propagator: III. The New vertex,''
  Class.\ Quant.\ Grav.\ \ {\bf 26}, 215001  (2009)
  [arXiv:0812.5018 [gr-qc]].

\bibitem{EuclideanAmplitude}
  J.~W.~Barrett, R.~J.~Dowdall, W.~J.~Fairbairn, H.~Gomes, F.~Hellmann,
  ``Asymptotic analysis of the EPRL four-simplex amplitude,''
  J.\ Math.\ Phys.\  {\bf 50}, 112504 (2009).
  [arXiv:0902.1170 [gr-qc]].

\bibitem{LorentzianAmplitude}
  J.~W.~Barrett, R.~J.~Dowdall, W.~J.~Fairbairn, F.~Hellmann and R.~Pereira,
  ``Lorentzian spin foam amplitudes: Graphical calculus and asymptotics,''
  Class.\ Quant.\ Grav.\  {\bf 27}, 165009 (2010)
  [arXiv:0907.2440 [gr-qc]].

\bibitem{Zakopane}
  C.~Rovelli,
  ``Zakopane lectures on loop gravity,''
  [arXiv:1102.3660 [gr-qc]].
  
\bibitem{LorentzReps} 
  V.~D.~Dao and V.~H.~Nguyen,
  ``On the theory of unitary representations of the sl(2c) group,''
  Acta Phys.\ Hung.\  {\bf 22}, 201 (1967).

\bibitem{VanishCosmology} 
  M.~Bojowald,
  ``The Semiclassical limit of loop quantum cosmology,''
  Class.\ Quant.\ Grav.\  {\bf 18}, L109 (2001)
  [gr-qc/0105113].

\bibitem{Vanish} 
  E.~Magliaro and C.~Perini,
  ``Emergence of gravity from spinfoams,''
  Europhys.\ Lett.\  {\bf 95}, 30007 (2011)
  [arXiv:1108.2258 [gr-qc]].

\bibitem{BH1}
 C.~Rovelli,
 ``Black hole entropy from loop quantum gravity,''
 Phys.\ Rev.\ Lett.\  {\bf 77}, 3288 (1996)
 [gr-qc/9603063].

\bibitem{BH2}
 A.~Ashtekar, J.~C.~Baez and K.~Krasnov,
 ``Quantum geometry of isolated horizons and black hole entropy,''
 Adv.\ Theor.\ Math.\ Phys.\  {\bf 4}, 1 (2000)
 [gr-qc/0005126].

\bibitem{BH3}
 J.~Engle, A.~Perez and K.~Noui,
 ``Black hole entropy and SU(2) Chern-Simons theory,''
 Phys.\ Rev.\ Lett.\  {\bf 105}, 031302 (2010)
 [arXiv:0905.3168 [gr-qc]].

\end{thebibliography}
\end{document}